\documentclass{ws-ijmpb}

\begin{document}

\markboth{SHIRO KAWABATA}
{2D MACROSCOPIC QUANTUM DYNAMICS IN YBCO JOSEPHSON JUNCTIONS}

%
\catchline{}{}{}{}{}
%

\title{TWO-DIMENSIONAL MACROSCOPIC QUANTUM DYNAMICS IN YBCO JOSEPHSON JUNCTIONS}

\author{SHIRO KAWABATA}

\address{Nanotechnology Research Institute (NRI), National Institute of Advanced Industrial Science and Technology (AIST), Tsukuba, Ibaraki, 305-8568, Japan, \\
Department of Microelectronics and Nanoscience (MC2), Chalmers University of Technology, S-41296 G\"oteborg, Sweden, and\\
CREST, Japan Science and Technology Corporation (JST), Kawaguchi, Saitama 332-0012, Japan\\
s-kawabata@aist.go.jp}

\author{TAKEO KATO}

\address{The Institute for Solid State Physics (ISSP), University of Tokyo, Kashiwa, Chiba, 277-8581, Japan\\
kato@issp.u-tokyo.ac.jp}

\author{FLORIANA LOMBARDI}

\address{Department of Microelectronics and Nanoscience (MC2), Chalmers University of Technology, S-41296 G\"oteborg, Sweden\\
floriana.lombardi@mc2.chalmers.se}

\author{THILO BAUCH}

\address{Department of Microelectronics and Nanoscience (MC2), Chalmers University of Technology, S-41296 G\"oteborg, Sweden\\
thilo.bauch@mc2.chalmers.se}



\maketitle

\begin{history}
\received{Day Month 2008}
\revised{Day Month 2008}
\end{history}

\begin{abstract}
We theoretically study classical thermal activation (TA) and macroscopic quantum tunneling (MQT) for a YBa${}_2$Cu${}_3$O${}_{7-\delta}$(YBCO) Josephson junction coupled with an LC circuit.
The TA and MQT escape rate are calculated by taking into account the two-dimensional nature of the classical and quantum phase dynamics. 
We find  that the MQT escape rate is largely suppressed by the coupling to the LC circuit.
On the other hand, this coupling leads to the slight reduction of the TA escape rate.
These results are relevant for the interpretation of a recent experiment on the MQT and TA phenomena in YBCO bi-epitaxial Josephson junctions.
\end{abstract}


\section{Introduction}

Macroscopic quantum tunneling (MQT) has become a focus of interest in physics and chemistry because it can provide a signature of quantum behavior in a macroscopic system.\cite{rf:Weiss}
Among several works on MQT, Josephson junctions have been intensively studied.\cite{rf:MQT4}
Heretofore, however, experimental and theoretical investigations of MQT have been focused on low-$T_c$ superconductor Josesphson junctions.

Renewed interest in MQT occurred after the recent experimental observations of MQT in high-$T_c$ superconductor Josephson junctions, e.g., YBa${}_2$Cu${}_3$O${}_{7-\delta}$ (YBCO) grain-boundary bi-epitaxial junctions\cite{rf:Bauch1,rf:Bauch2} and Bi${}_2$Sr${}_2$CaCu${}_2$O${}_{8+\delta}$ intrinsic junctions.\cite{rf:Inomata1,rf:Jin,rf:Matsumoto,rf:Li,rf:YurgensKadowaki,rf:Kashiwaya1,rf:Kashiwaya2}
Such fascinating results open up the possibility for realizing high-$T_c$ quantum bits (qubits).
Over the past year much progress has been achieved for making theory of the MQT\cite{rf:Kawabata1,rf:Kawabata2,rf:Kawabata3,rf:Machida,rf:Fistul,rf:Nori1,rf:Kawabata4,rf:Yokoyama} and macroscopic quantum coherence\cite{rf:Khveshchenko,rf:Umeki} in high-$T_c$ junctions.

\begin{figure}[b]
\begin{center}
\includegraphics[width=12.5cm]{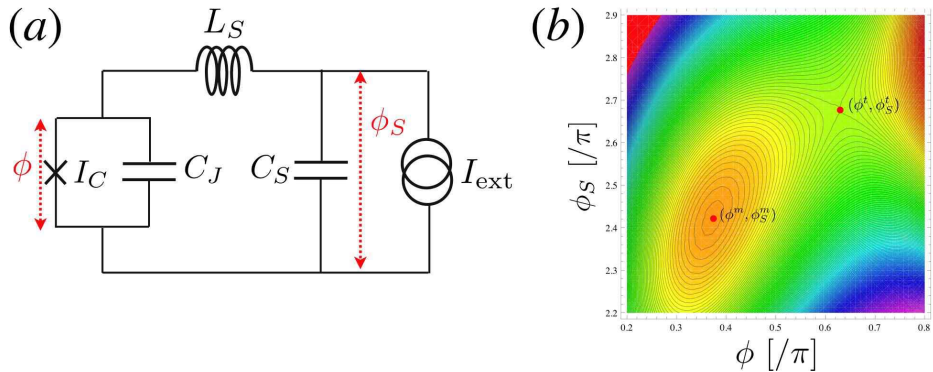}
\caption{(a) The extended circuit model for a bi-epitaxial YBCO junction, including the bias current $I_\mathrm{ext}$, the stray capacitance $C_S$ and kinetic inductance $L_S$.
$\phi$ and $\phi_S$ are the phase difference across the Josephson junction and the stray capacitance $C_S$, respectively.
(b) The two-dimensional potential profile $U(\phi,\phi_S)$ of (a) for $\gamma=I_\mathrm{ext}/I_C=0.92$ and $\eta=L_S/L_{J0}=7$.
}
\end{center}
\label{fig:1}
\end{figure}

In YBCO grain boundary junctions which was used in MQT experiments,\cite{rf:Bauch1,rf:Bauch2} it was found that the stray capacitance $C_S$ of the electrodes, due to the large dielectric constant $\epsilon$ of the STO substrate at low temperature ($\epsilon>10000$), and the inductance $L_S$, due to the large London penetration depth in $c$-axis and/or Josephson coupling between CuO${}_2$ planes in one of the electrodes have large influence on the macroscopic dynamics and can be taken into account by an extended circuit model (see Fig.1(a)).
In Fig. 1, $\phi$ is the phase difference across the Josephson junction and $\phi_s= (2 \pi/\Phi_0) I_S L_S + \phi$ is the phase difference across the capacitor $C_S$, where $I_S$ is the current through the inductor.
As will be mentioned later, the addition of the LC circuit results in a two-dimensional potential $U(\phi,\phi_S)$ which makes the dynamics much more complex than for an ordinary single junction (see Fig.1 (b)).
It was found that, in the microwave-assisted MQT experiment,\cite{rf:Bauch2} the bias current $I_\mathrm{ext}$ dependence of the Josephson plasma frequency $\omega_p$ is quantitatively explained by this model.\cite{rf:Lombardi,rf:Rotoli}
However, the validity of the extended circuit model for thermal activation (TA)  and MQT escape processes have not yet been explored.
In this paper, we investigate the validity of the extended circuit model for both the TA and MQT escape processes and discuss the role of the two-dimensional nature on the macroscopic quantum dynamics in the YBCO bi-epitaxial junction.\cite{rf:Kawabata5,rf:Kawabata6}

\section{Model and Lagrangian}
In this section we derive the Lagrangian for  the Josephson junction coupled to the LC circuit and discuss the two-dimensional potential structure of this model.
The Hamiltonian of the circuit [Fig. 1(a)] can be written as
\begin{eqnarray}
{\cal H}
&=&
\frac{Q_J^2}{2 C_J}
+
\frac{Q_S^2}{2 C_S}
-E_J \cos \phi
+
\left( \frac{\Phi_0}{ 2 \pi}\right)^2 
\frac{\left( \phi-\phi_S\right)^2}{2 L_S} 
-
\left( \frac{\Phi_0}{ 2 \pi} \right)
 \phi_S I_\mathrm{ext} 
 ,
\label{eq:1-1}
\end{eqnarray}
where  $Q_{J}=C_{J}(\Phi_0/2 \pi) (d \phi /d t)$ and  $Q_{S}=C_{S} (\Phi_0/2 \pi) $ $(d \phi_S /d t)$ are the charge on the junction and the stray capacitor $C_S$, respectively, $E_J=(\hbar/2 e) I_C$ is the Josephson coupling energy, and $\Phi_0=h/2e$ is the flux quantum.

The Lagrangian is given by 
\begin{eqnarray}
{\cal L}
\! = 
\frac{C_J}{2}
\left( 
\frac{\Phi_0}{2 \pi}
\frac{\partial \phi}{\partial t}
\right)^2
+
\frac{C_S}{2}
\left( 
\frac{\Phi_0}{2 \pi}
\frac{\partial \phi_S}{\partial t}
\right)^2
-
U(\phi,\phi_S)
,
\label{eq:1-2}
\end{eqnarray}
\begin{eqnarray}
U(\phi,\phi_S)
&=&
E_J \left[
- \cos \phi + \frac{ \left(\phi-\phi_S \right)^2}{2 \eta}
-
\gamma \phi_S
\right]
,
\label{eq:1-3}
\end{eqnarray}
where $\gamma=I_\mathrm{ext}/I_C$ and $\eta \equiv 2 \pi I_C L_S / \Phi_0= L_S/L_{J0}$ with $L_{J0} = \Phi_0/2 \pi I_C$ being the zero-bias Josephson inductance.
The Lagrangian ${\cal L}$ describes the quantum dynamics of a fictive particle moving in a two-dimensional tilted washboard potential $U(\phi,\phi_S)$.
Therefore the escape paths traverse a two-dimensional land scape, so the macroscopic dynamics in this model becomes more complicated than that in the simple one-dimensional model.

The two-dimensional potential profile $U(\phi,\phi_S)$ is shown in Fig. 1(b).
The mean slope along the $\phi_S$ direction is proportional to the bias current $I_\mathrm{ext}$, and the mean curvature in the diagonal direction ($\phi=\phi_S)$ is due to the inductive coupling between the Josephson junction and capacitance $C_S$ characterized by $\eta$.
The local minimum point $(\phi^m,\phi_S^m)$ and the saddle point $(\phi^t,\phi_S^t)$ is given from Eq. (\ref{eq:1-3}) as 
\begin{eqnarray}
\left( \phi^m,\phi_S^m \right)&=&
\left( \sin^{-1} \gamma, \eta \gamma +   \sin^{-1} \gamma \right)
,
\label{eq:1-4}
\\
\left(\phi^t,\phi_S^t \right)&=&
\left( \pi - \sin^{-1} \gamma,  \pi  -   \sin^{-1} \gamma + \eta \gamma \right)
,
\label{eq:1-5}
\end{eqnarray}
Then the potential barrier height $V_0^\mathrm{2D}$ is given by
\begin{eqnarray}
V_0^\mathrm{2D} &\equiv& U \left(\phi^t,\phi_S^t \right) - U \left( \phi^m,\phi_S^m \right) 
=
E_J \left(
2 \gamma \sin^{-1} \gamma - \pi \gamma + 2 \sqrt{1-\gamma^2}
\right)
,
\label{eq:1-6}
\end{eqnarray}
and is a decreasing function of $\gamma$.
Importantly, the potential barrier height $V_0^\mathrm{2D}$ does not depend on the LC circuit parameters $L_S$ and $C_S$.\cite{rf:Rotoli}
Moreover, the expression of $V_0^\mathrm{2D}$ is identical with the barrier height for usual single Josephson junctions, i.e., one-dimensional model.\cite{rf:MQT4}

Next we will derive an approximate expression of $U(\phi,\phi_S)$ for $\gamma \lesssim 1$. 
By introducing the new coordinate $(x,y) = (\phi - \phi^m, \phi_S - \phi_S^m)$ and assuming $\gamma \lesssim 1$, we can rewrite the Lagrangian as 
\begin{eqnarray}
{\cal L}
&=&
\frac{M}{2}
\left(
\frac{\partial x}{\partial t}
\right)^2
+
\frac{m}{2}
\left(
\frac{\partial y}{\partial t}
\right)^2
-
U(x,y)
,
\label{eq:1-7}
\\
U(x,y)
&=&
U_\mathrm{1D} (x)
+
E_J \frac{(x-y)^2}{2 \eta}
,
\label{eq:1-8}
\end{eqnarray}
where $M=C_J (\Phi_0 / 2 \pi)^2 $ and $m= C_S (\Phi_0 / 2 \pi)^2$.
Here  $U_\mathrm{1D} (x) \approx (1/2) M \omega_p^2   (x^2 -x^3/x_1)$ is the potential of the Josephson junction without the LC circuit, where $x_1=3\sqrt{1- \gamma^2}$ and  $\omega_p = \omega_{p0} (1- \gamma^2)^{1/4}$ is the Josephson plasma frequency with $\omega_{p0} = \sqrt{2 \pi I_C / \Phi_0 C_J}$ being the zero-bias plasma frequency.

\section{Effective Action}
In this section we will derive an effective action from the Lagrangian (\ref{eq:1-7}).
By using the functional integral method,\cite{rf:Weiss,rf:Zaikin} the partition function ${\cal Z}$ of the system can be written as
\begin{eqnarray}
{\cal Z} 
=
\int  {\cal D} x (\tau) 
\int  {\cal D} y (\tau) 
\exp
\left[
-\frac{1}{\hbar} \int_0^{\hbar \beta} d \tau {\cal L} [x,y] 
\right]
,
\label{eq:2-1}
\end{eqnarray}
where
\begin{eqnarray}
 {\cal L} [x,y] 
=
\frac{M}{2} \dot{x}^2  +\frac{m}{2}  \dot{y}^2  + U(x,y)
,
\end{eqnarray}
is the Euclidean Lagrangian and $\beta=1/ k_B T$.
The Lagrangian is a quadratic function of $y$ and the coupling term between $x$ and $y$ is linear, so the functional integral over variable $y$ can be performed explicitly by use of the Feynman-Vernon influence functional technique.\cite{rf:Weiss}
Then the partition function is reduced to a single functional integral over $x$, i. e., 
\begin{eqnarray}
{\cal Z} = \int {\cal D} x(\tau) \exp \left(
- \frac{{\cal S}_\mathrm{eff}^\mathrm{2D} [x] }{\hbar}
\right),
\end{eqnarray}
 where the effective action is given by 
\begin{eqnarray}
{\cal S}_\mathrm{eff}^\mathrm{2D}[x]&=&{\cal S}_\mathrm{eff}^\mathrm{1D}[x]+{\cal S}_\mathrm{eff}^\mathrm{ret}[x]
\\
{\cal S}_\mathrm{eff}^\mathrm{1D}[x]
&=&
\int_0^{\hbar \beta}
d \tau \left[
\frac{1}{2} M \dot{x}^2 + U_\mathrm{1D}  (x)
\right] 
,
\label{eq:2-2}
\\
{\cal S}_\mathrm{eff}^\mathrm{ret}[x]
&=&
\frac{1}{4}
\int_{0}^{\hbar \beta} d \tau
\int_{0}^{\hbar \beta} d \tau'
\left[ x(\tau) - x(\tau') \right]^2  K (\tau-\tau') 
\!
.
\label{eq:2-3}
\end{eqnarray}
Thus the dynamics of the phase difference in a two-dimensional potential $U(\phi,\phi_S)$ can be mapped into one in an one-dimensional model.
Note that due to the coupling between the junction and the LC circuit, the effective action ${\cal S}_\mathrm{eff}^\mathrm{2D}[x]$ contains a kind of $dissipation$ action ${\cal S}_\mathrm{eff}^\mathrm{ret}[x]$ in a sense that the retarded (or nonlocal) effect exists.
The nonlocal kernel $K(\tau)$ in Eq. (\ref{eq:2-3}) is defined by
\begin{eqnarray}
K(\tau)
&=&
\frac{1}{2} m \omega_{LC}^3
\frac{
      \cosh \left[  \omega_{LC} \left(  \frac{\hbar \beta}{2} - \left| \tau \right| \right)\right]
      }
      {
      \sinh \left[  \frac{ \hbar \beta \omega_{LC} }{2}  \right]
      }
       .
       \label{eq:2-4}
\end{eqnarray}
In the next section, we will calculate the thermal and quantum escape rate from the effective action ${\cal S}_\mathrm{eff}^\mathrm{2D}$.

\section{Thermal Activation Process}

The TA escape rate well above the crossover temperature $T_\mathrm{co}$ is given by\cite{rf:Weiss}  
\begin{eqnarray}
\Gamma_\mathrm{TA}^\mathrm{2D}
& =&
\frac{\omega_R}{2 \pi}
c_\mathrm{qm}^\mathrm{2D}
\exp \left(  - \frac{V_0^\mathrm{2D}}{k_B T} \right)
,
\label{eq:4-1}
\\
c_\mathrm{qm}^\mathrm{2D}
& =&
 \prod _{n=1}^\infty 
 \frac{\omega_n^2 + \omega_p^2 + \omega_n \hat{\gamma} (\omega_n) }
 {\omega_n^2 - \omega_p^2 + \omega_n \hat{\gamma} (\omega_n) } 
 ,
\label{eq:4-2}
 \end{eqnarray}
where $c_\mathrm{qm}^\mathrm{2D}$ is the quantum mechanical enhancement factor resulting form stable fluctuation modes, $\omega_n$ is the Matsubara frequency, and $\hat{\gamma}(\omega_n)=(C_S/C_J) |\omega_n| \omega_{LC}^2/(\omega_n^2 +\omega_{LC}^2)$ is the Fourier transform of the memory-friction kernel.
The potential barrier height is not changed even in the presence of the LC circuit, so $V_0^\mathrm{2D}=V_0^\mathrm{1D}$.
Thus the coupling to the LC circuit modifies only the prefactor of $\Gamma_\mathrm{TA}$.

For the non-adiabatic ($\omega_p \gg \omega_\mathrm{LC}$) cases which can be applicable to actual experiment,\cite{rf:Bauch1} the quantum enhancement factor $c_\mathrm{qm}^\mathrm{2D}$ (\ref{eq:4-2}) is approximately given by 
\begin{eqnarray}
c_\mathrm{qm}^\mathrm{2D}
&\approx&
   \frac{\sinh \left(  \frac{\hbar \beta \omega_p}{2 \sqrt{1+  \frac{C_S}{C_J} }} \right)}
          {\sin \left(  \frac{\hbar \beta \omega_p}{2 \sqrt{1+ \frac{C_S}{C_J} }} \right)}
.
\label{eq:4-4}
\end{eqnarray}
In the case of $C_S/C_J \ll 1$, the quantum enhancement factor $c_\mathrm{qm}^\mathrm{2D}$ coincides with the result without retardation effects, i.e., $c_\mathrm{qm}^\mathrm{1D}=   \sinh \left(  \hbar \beta \omega_p  / 2  \right) /    \sin \left(  \hbar \beta \omega_p / 2  \right)$.\cite{rf:Weiss}
Therefore, the influence of the coupling to the LC circuit on the thermal activation process is quite weak, so the system behaves as one-dimensional systems well above the crossover temperature $T_\mathrm{co}$.
In Sec. 6 we will numerically compare theoretical results with experimental data in the TA regime.

\section{Macroscopic Quantum Tunneling Process}
The MQT escape rate at zero temperature\cite{rf:Weiss} is given by 
\begin{eqnarray}
\Gamma_\mathrm{MQT}^\mathrm{2D}= \lim_{\beta \to \infty} \left(  \frac{2}{\beta}  \right)
\mathrm{Im} \ln {\cal Z}
\end{eqnarray}
By use of the bounce techniques, the MQT escape rate $\Gamma_\mathrm{MQT}^\mathrm{2D}$ is perturbatively determined by 
\begin{eqnarray}
\Gamma_\mathrm{MQT}^\mathrm{2D}(T=0)
=
\!
\frac{\omega_p}{2 \pi} \sqrt{120 \pi\left(  B_\mathrm{1D}+ B_\mathrm{ret} \right)  } e^{   - B_\mathrm{1D} - B_\mathrm{ret}  }
,
\label{eq:5-5}
 \end{eqnarray}
where $B_\mathrm{1D} = {\cal S}_\mathrm{eff}^\mathrm{1D} [x_B]/\hbar=36 V_0^\mathrm{1D} / 5 \hbar \omega_p$ and $B_\mathrm{ret}= {\cal S}_\mathrm{eff}^\mathrm{ret} [x_B]/\hbar$ are the bounce exponents, that are the value of the actions evaluated along the bounce trajectory $x_B(\tau)$.
In the zero temperature and the non-adiabatic limit, the bounce action ${\cal S}_\mathrm{eff}^\mathrm{ret}[x_B]$ is analytically given by 
\begin{eqnarray}
{\cal S}_\mathrm{eff}^\mathrm{ret}[x_B]
\approx
\frac{4}{3} m \left( \frac{\omega_{LC}}{\omega_p} \right)^2 \omega_p  x_1^2.
\label{eq:5-7}
\end{eqnarray}
%
%
%
Thus the total bounce exponent is given by 
\begin{eqnarray}
B_\mathrm{1D} + B_\mathrm{diss} 
&=&
\frac{8}{15 \hbar}
(M +\delta M) \omega_p x_1^2
,
\label{eq:5-8}
\end{eqnarray}
where 
\begin{eqnarray}
\frac{\delta M}{M}
&\approx&
\frac{5}{2}   \frac{L_J}{L_S}  
,
\label{eq:5-9}
\end{eqnarray}
is the retardation correction to the mass $M$.
Thus, due to the two-dimensional nature of the phase dynamics, the bounce exponent is increased with compared to one-dimensional cases.

By substituting Eq. (\ref{eq:5-8}) into Eq. (\ref{eq:5-5}), we finally get the zero-temperature MQT escape rate as
\begin{eqnarray}
\Gamma_\mathrm{MQT}^\mathrm{2D}  (T=0)
&=&
\frac{\omega_p}{2 \pi} \sqrt{864 \pi \frac{V_0^\mathrm{1D}}{\hbar \omega_p}  \left( 1+\frac{\delta M}{M} \right) }
 \exp \left[ 
 - \frac{36}{5} \frac{V_0^\mathrm{1D}}{\hbar \omega_p}  \left( 1+\frac{\delta M}{M} \right) 
 \right]
 .
 \label{eq:5-10}
 \end{eqnarray}
Therefore the coupling to the LC circuit effectively increases the barrier potential, $i.e.,$ $V_0^\mathrm{1D} \to V_0^\mathrm{1D} (1+\delta M/M)$. 
In contrast to the TA escape rate, the coupling to the LC circuit gives rise to reduce the MQT escape rate $\Gamma_\mathrm{MQT}^\mathrm{2D}$ considerably.
Therefore the two-dimensional nature has large influence on the MQT escape process at the low temperature regime.

\section{Comparison with Experiment}
In order to check the validity of the extended circuit model\cite{rf:Bauch2,rf:Lombardi,rf:Rotoli} for the TA and MQT escape process, we try to compare our result with the experimental data\cite{rf:Bauch1} of the switching current distribution at the high-temperature TA and the low temperature MQT regimes.
In these estimation, we will use $I_C=1.4$ $\mu$A, $L_S=1.7$ nH and $C_S=1.6$ pF which have been directly determined from the TA and the microwave-assisted MQT experiments.\cite{rf:Bauch1,rf:Bauch2}
We have numerically calculated the switching current distribution $P(\gamma)$ which is related to the escape rate $\Gamma$ as\cite{rf:Voss,rf:Garg} 
\begin{eqnarray}
P(\gamma)=\frac{1}{v}
 \Gamma (\gamma) \exp
\left[
- \frac{1}{v}
\int_0^{\gamma} \Gamma (\gamma') d \gamma'
\right]
,
 \label{eq:6-1}
\end{eqnarray}
where $v \equiv \left| d \eta / d t \right| $ is the sweep rate of the external bias current.
In the actual experiment,\cite{rf:Bauch1} the temperature dependence of the full width at half maximum (HMFW) $\sigma$ of $P(\gamma)$ is measured.

\begin{figure}[tb]
\begin{center}
\includegraphics[width=12cm]{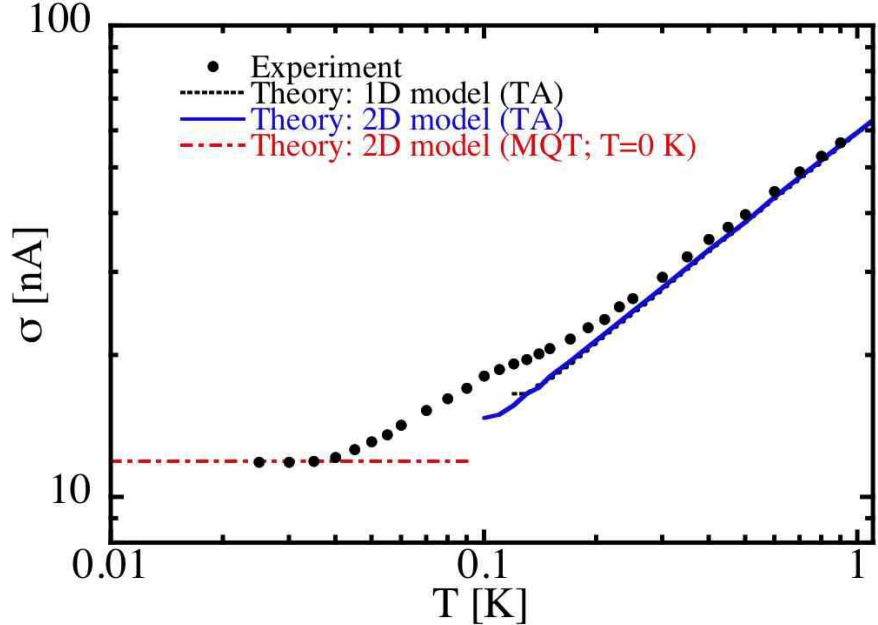}
\caption{The temperature $T$ dependence of the full width at half maximum $\sigma$ of the switching current distribution $P(\gamma)$.
Both the one- (dotted black) and two-dimensional model (blue solid) give almost same result above the crossover temperature.
The calculated $\sigma$ at $T=0$ K from the zero-temperature escape rate is shown by dashed-dotted (red) line.
Experimental data of $\sigma$ (black circles) for a YBCO bi-epitaxial Josephson junction is also plotted.
}
\end{center}
\label{fig:5}
\end{figure}

Firstly we will investigate the TA regime.
In Fig. 2 we show the temperature dependence of $\sigma$ in the TA escape regime (blue solid and black dotted lines).
Both the one- and two-dimensional model give good agreement with the experimental data (black circles) well above the crossover temperature ($T_\mathrm{co}^\mathrm{exp} \sim 0.04$ K).
Therefore, in the TA regime, the system can be served as an one-dimensional model without the LC circuit.

In the MQT regime, the measured saturated-value of $\sigma$  is found to be 11.9 nA.\cite{rf:Bauch1}
In order to numerically calculate $\sigma$, we need the information about $I_C$, $C_J$, $L_S$, and $C_S$.
The values of $I_C=1.4$ $\mu$A, $L_S=1.7$ nH and $C_S=1.6$ pF have been directly determined from the experiments.\cite{rf:Bauch1,rf:Bauch2}
Therefore the only fitting parameter is $C_J$.
From the numerical estimation of $\sigma$, we found that $C_J =0.22$ pF gives good agreement with the experimental value of $\sigma$.
The obtained value of $C_J$ is consistent with the estimated value  $C_J \approx 0.16$ pF based on the geometry of the junction.\cite{rf:Rotoli}
Therefore, we can conclude that the extended circuit model can quantitatively explain the MQT experiment\cite{rf:Bauch1} in the YBCO bi-epitaxial junction.

\section{Conclusions}

In the present work, the TA and the MQT escape process of the YBCO Josephson junction coupled to the LC circuit has been analyzed by taking into account the two-dimensional nature of the phase dynamics.
Based on the Feynman-Vernon approach,  the effective one-dimensional effective action is derived by integrating out the degree of freedom for the LC circuit.
We found that the coupling to the LC circuit gives negligible reduction for TA escape rate.
On the other hand, we also found that the MQT escape rate is considerably reduced due to the coupling between the junction and  the LC circuit.
These results are consistent with experimental  result of a YBCO bi-epitaxial Josephson junction.\cite{rf:Bauch1}
In our model, we have assumed that the Josephson current-phase relation is given by $I_J=I_C \sin \phi$.
Therefore our theory can also be applicable to low-$T_c$ josephson junctions coupled to the artificial LC circuit.
We expect that our prediction will be confirmed by use of not only high-$T_c$ but also the low-$T_c$ junctions experimentally.

\section*{Acknowledgements}

We would like to thank J. Ankerhold, A. Barone, M. Fogelstr\"om, A. A. Golubov,  G. Johansson, J. R. Kirtley, J. P. Pekola, G. Rotoli, V. S. Shumeiko, and F. Tafuri  for useful discussions. 
One of the authors (S. K.) would like to thank the Applied Quantum Physics Laboratory at the Chalmers University of Technology, for its hospitality during the course of this work.
This work was supported by the NanoNed Program under Project No. TCS. 7029, JST-CREST, and the JSPS-RSAS Scientist Exchange Program.

\end{document}